# Superactivation of Quantum Channels is Limited by the Quantum Relative Entropy Function

Laszlo Gyongyosi*, Sandor Imre

Quantum Technologies Laboratory, Department of Telecommunications

Budapest University of Technology and Economics

Budapest, Hungary

*gyongyosi@hit.bme.hu

**In this work we prove that the possibility of superactivation of quantum channel capacities is determined by the mathematical properties of the quantum relative entropy function. Before our work this fundamental and purely mathematical connection between the quantum relative entropy function and the superactivation effect was completely unrevealed. We demonstrate the results for the quantum capacity; however the proposed theorems and connections hold for all other channel capacities of quantum channels for which the superactivation is possible.**

## 1 Introduction

In the first decade of the 21st century, many revolutionary properties of quantum channels were discovered. These phenomena are purely quantum mechanical and completely unimaginable in classical systems. Recently, one of the most important discoveries in Quantum Shannon Theory was the possibility of transmitting quantum information over zero-capacity quantum channels. The superactivation of quantum channels is an extreme violation of the additivity [1-3], [26] of quantum channels. This effect makes possible the communication over zero-capacity quantum channels. The superactivation effect was discovered by Smith and Yard in 2008 [4], who with Smolin demonstrated experimentally that this effect works for the quantum capacity [5]. Later, these results were extended to the classical zero-error capacity by Duan [6], and Cubitt et al. [7] and to the quantum zero-error capacity by Cubitt and Smith [8]. An algorithmic solution to the problem was developed in [21]. The impossibility of superactivation of classical zero-error capacity of qubit channels was shown in [28]. Currently, we have no theoretical background for describing all possible combinations of superactive zero-capacity channels; hence, there may be many other possible combinations [22-27].

In this paper we prove that the problem of superactivation is rooted in information geometric issues and there is a strict connection between the mathematical properties of the quantum relative entropy function and the possibility of superactivation. As we have discovered, the set of superactive channel combinations is limited and determined by the quantum relative entropy function. The results are illustrated with the $Q(\mathcal{N}_1 \otimes \mathcal{N}_2)$ quantum capacity of the joint structure, $\mathcal{N}_1 \otimes \mathcal{N}_2$. However, the proposed theorems hold for all channel capacities of the joint channel $\mathcal{N}_1 \otimes \mathcal{N}_2$ for which the superactivation is possible. These capacities are the single-use and asymptotic quantum capacities $Q^{(1)}(\mathcal{N}_1 \otimes \mathcal{N}_2)$, $Q(\mathcal{N}_1 \otimes \mathcal{N}_2)$, the zero-error classical capacities $C_0^{(1)}(\mathcal{N}_1 \otimes \mathcal{N}_2)$, $C_0(\mathcal{N}_1 \otimes \mathcal{N}_2)$ and the zero-error quantum capacities $Q_0^{(1)}(\mathcal{N}_1 \otimes \mathcal{N}_2)$, $Q_0(\mathcal{N}_1 \otimes \mathcal{N}_2)$.

This paper is organized as follows. In Section 2 we reveal some important connections between the quantum capacity, Holevo information and the quantum relative entropy function. In Section 3 we discuss the theorems and proofs. Finally, in Section 4 we conclude the results.

## 2 Quantum Capacity of a Quantum Channel

The classical and the quantum capacities of quantum channels are described by the Holevo-Schumacher-Westmoreland (HSW) [12-13] and the Lloyd-Shor-Devetak (LSD) [9-11] theorems. In case of the quantum capacity $Q(\mathcal{N})$, the correlation measure is the *quantum coherent information* function. The single-use quantum capacity of quantum channel $\mathcal{N}$ is the maximization of the $I_{coh}$ quantum coherent information:

$$Q^{(1)}(\mathcal{N}) = \max_{\text{all } p_i, \rho_i} I_{coh} \tag{1}$$

The $I_{coh}(\rho_A : \mathcal{N}(\rho_A))$ quantum coherent information can be expressed as

$$\begin{aligned} I_{coh}(\rho_A : \mathcal{N}(\rho_A)) &= S(\mathcal{N}(\rho_A)) - S_E(\rho_A : \mathcal{N}(\rho_A)) \\ &= S(\rho_B) - S(\rho_E), \end{aligned} \tag{2}$$

where $S(\rho) = -Tr(\rho \log(\rho))$ is the von Neumann entropy and $S_E(\rho_A : \mathcal{N}(\rho_A))$ is the entropy exchange.

## 2.1 Connection between Quantum Coherent Information and Holevo Information

In the proof we exploit a *connection*[*] between the Holevo information and the quantum coherent information. As it has been shown by Schumacher and Westmoreland [17], the quantum coherent information also can be expressed with the help of Holevo information, as follows

$$I_{coh}(\rho_A : \mathcal{N}(\rho_A)) = (\mathcal{X}_{AB} - \mathcal{X}_{AE}), \tag{3}$$

where

$$\mathcal{X}_{AB} = S(\mathcal{N}_{AB}(\rho_{AB})) - \sum_i p_i S(\mathcal{N}_{AB}(\rho_i)) \tag{4}$$

and

$$\mathcal{X}_{AE} = S(\mathcal{N}_{AE}(\rho_{AE})) - \sum_i p_i S(\mathcal{N}_{AE}(\rho_i)) \tag{5}$$

measure the Holevo quantities between Alice and Bob, and between Alice and environment E, where $\rho_{AB} = \sum_i p_i \rho_i$ and $\rho_{AE} = \sum_i p_i \rho_i$ are the average states. As follows, the *single-use* quantum capacity $Q^{(1)}(\mathcal{N})$ can be expressed as

$$Q^{(1)}(\mathcal{N}) = \max_{\text{all } p_i, \rho_i} (\mathcal{X}_{AB} - \mathcal{X}_{AE})$$

$$= \max_{\text{all } p_i, \rho_i} S\left(\mathcal{N}_{AB}\left[\sum_{i=1}^{n} p_i(\rho_i)\right]\right) - \sum_{i=1}^{n} p_i S(\mathcal{N}_{AB}(\rho_i)) \quad (6)$$

$$- S\left(\mathcal{N}_{AE}\left[\sum_{i=1}^{n} p_i(\rho_i)\right]\right) + \sum_{i=1}^{n} p_i S(\mathcal{N}_{AE}(\rho_i)),$$

where $\mathcal{N}(\rho_i)$ represents the $i$-th output density matrix obtained from the quantum channel input density matrix $\rho_i$. The *asymptotic* quantum capacity $Q(\mathcal{N})$ can be expressed by

$$Q(\mathcal{N}) = \lim_{n \to \infty} \frac{1}{n} Q^{(1)}(\mathcal{N}^{\otimes n}) = \lim_{n \to \infty} \frac{1}{n} \max_{\text{all } p_i, \rho_i} I_{coh}(\rho_A : \mathcal{N}^{\otimes n}(\rho_A))$$

$$= \lim_{n \to \infty} \frac{1}{n} \max_{\text{all } p_i, \rho_i} (\mathcal{X}_{AB} - \mathcal{X}_{AE})^{\otimes n}. \quad (7)$$

---

[*] *This connection is a rather surprising but not well known result in Quantum Information Theory, for the details see the proof of Eq. 70 in Ref. [17].*

As summarize, the quantum capacity $Q(\mathcal{N})$ of a quantum channel $\mathcal{N}$ can be defined by $\mathcal{X}_{AB}$, the *Holevo quantity* of Bob's output and by $\mathcal{X}_{AE}$, the information leaked to the environment during the transmission.

## 2.2 Connection between Holevo Information and Quantum Relative Entropy Function

The quantum relative entropic distance between quantum states $\rho$ and $\sigma$ is defined by the *quantum relative entropy* function $D(\cdot\|\cdot)$ as

$$D(\rho\|\sigma) = Tr(\rho \log(\rho)) - Tr(\rho \log(\sigma))$$
$$= Tr[\rho(\log(\rho) - \log(\sigma))]. \quad (8)$$

The Holevo quantity can be expressed by the quantum relative entropy function as [18-20], [14-17]

$$\chi = D(\rho_k\|\sigma), \quad (9)$$

where $\rho_k$ denotes an *optimal* (for which the Holevo quantity will be maximal) channel *output state* and $\sigma = \sum p_k \rho_k$ is the mixture of the optimal output states [16]. The Holevo information $\mathcal{X}$ can be derived in terms of the quantum relative entropy in the following way [14-17]

$$\begin{aligned}
\sum_k p_k D(\rho_k \| \sigma) &= \sum_k \left( p_k Tr(\rho_k \log(\rho_k)) - p_k Tr(\rho_k \log(\sigma)) \right) \\
&= \sum_k \left( p_k Tr(\rho_k \log(\rho_k)) \right) - Tr\left[ \sum_k (p_k \rho_k \log(\sigma)) \right] \\
&= \sum_k \left( p_k Tr(\rho_k \log(\rho_k)) \right) - Tr(\sigma \log(\sigma)) \\
&= S(\sigma) - \sum_k p_k S(\rho_k) = \mathcal{X}.
\end{aligned} \quad (10)$$

We express the Holevo information between Alice and Bob as

$$\mathcal{X}_{AB} = S\left(\mathcal{N}_{AB}\left[\sum_{i=1}^{n} p_i \rho_i\right]\right) - \sum_{i=1}^{n} p_i S(\mathcal{N}_{AB}(\rho_i)) = D(\rho_k^{AB} \| \sigma^{AB}). \quad (11)$$

The second quantity measures the Holevo information which is leaked to the environment during the transmission as

$$\mathcal{X}_{AE} = S\left(\mathcal{N}_{AE}\left[\sum_{i=1}^{n} p_i \rho_i\right]\right) - \sum_{i=1}^{n} p_i S(\mathcal{N}_{AE}(\rho_i)) = D(\rho_k^{AE} \| \sigma^{AE}). \quad (12)$$

Using the resulting quantum relative entropy function and the Lloyd-Shor-Devetak (LSD) theorem [9-11], the asymptotic LSD capacity $Q(\mathcal{N})$ can be expressed with as follows

$$\begin{aligned}
Q(\mathcal{N}) &= \lim_{n \to \infty} \frac{1}{n} Q^{(1)}(\mathcal{N}^{\otimes n}) \\
&= \lim_{n \to \infty} \frac{1}{n} \max_{p_1,\dots,p_n,\rho_1,\dots,\rho_n} I_{coh}(\rho_A : \mathcal{N}^{\otimes n}(\rho_A)) \\
&= \lim_{n \to \infty} \frac{1}{n} \max_{p_1,\dots,p_n,\rho_1,\dots,\rho_n} (\mathcal{X}_{AB} - \mathcal{X}_{AE})^{\otimes n} \\
&= \lim_{n \to \infty} \frac{1}{n} \max_{p_1,\dots,p_n,\rho_1,\dots,\rho_n} S\left(\mathcal{N}_{AB}^{\otimes n}\left[\sum_{i=1}^{n} p_i \rho_i\right]\right) - \sum_{i=1}^{n} p_i S(\mathcal{N}_{AB}^{\otimes n}(\rho_i)) \\
&\quad - S\left(\mathcal{N}_{AE}^{\otimes n}\left[\sum_{i=1}^{n} p_i \rho_i\right]\right) + \sum_{i=1}^{n} p_i S(\mathcal{N}_{AE}^{\otimes n}(\rho_i)) \\
&= \lim_{n \to \infty} \frac{1}{n} \sum_n \left( \min_{\sigma_{1\dots n}} \max_{\rho_{1\dots n}} D(\rho_k^{AB} \| \sigma^{AB}) - \min_{\sigma_{1\dots n}} \max_{\rho_{1\dots n}} D(\rho_k^{AE} \| \sigma^{AE}) \right) \\
&= \lim_{n \to \infty} \frac{1}{n} \sum_n \left( \min_{\sigma_{1\dots n}} \max_{\rho_{1\dots n}} D(\rho_k^{AB-AE} \| \sigma^{AB-AE}) \right),
\end{aligned} \quad (13)$$

where $\mathcal{X}_{AB}$ is the *Holevo quantity* of Bob's output, $\mathcal{X}_{AE}$ is the information leaked to the environment during the transmission, $\rho_k^{AB}$ is Bob's optimal output state, $\rho_k^{AE}$ is the environment's optimal state, $\sigma^{AB}$ is Bob's optimal output average state, $\sigma^{AE}$ is the environment's average state, while $\rho_k^{AB-AE}$ is the final optimal output channel state and $\sigma^{AB-AE}$ is the final output average state. The term *AB-AE* denotes the information which is transmitted from Alice to Bob minus the information which is leaked to the environment during the transmission. For joint structure $\mathcal{N}_{12} = \mathcal{N}_1 \otimes \mathcal{N}_2$ the single-use joint quantum capacity can be expressed by the $D(\cdot \| \cdot)$ quantum relative entropy function as

$$\begin{aligned} Q^{(1)}(\mathcal{N}_1 \otimes \mathcal{N}_2) &= \min_\sigma \max_\rho \left( D\left(\rho_{12}^{AB} \| \sigma_{12}^{AB}\right) - D\left(\rho_{12}^{AE} \| \sigma_{12}^{AE}\right) \right) \\ &= \min_\sigma \max_\rho D\left(\rho_{12}^{AB-AE} \| \sigma_{12}^{AB-AE}\right), \end{aligned} \quad (14)$$

$\rho_{12}^{AB}$ is the optimal output state of joint channel $\mathcal{N}_{12}$, and $\sigma_{12}^{AB}$ is the average state of joint channel $\mathcal{N}_{12}$ between Alice and Bob. The term $E$ denotes the environment, and $AE$ is the channel between Alice and the environment with the optimal state $\rho_{12}^{AE}$, and average state $\sigma_{12}^{AE}$.

The final optimal output channel state is depicted by $\rho_{12}^{AB-AE}$, while $\sigma_{12}^{AB-AE}$ is the final output average state of the channel between Alice and the environment. $Q^{(1)}(\mathcal{N}_{12}) > 0$ only if the $\mathcal{N}_{12} = \mathcal{N}_1 \otimes \mathcal{N}_2$ joint structure is superactive, otherwise $Q^{(1)}(\mathcal{N}_1) = Q^{(1)}(\mathcal{N}_2) = Q^{(1)}(\mathcal{N}_{12}) = 0$.

## 3 Theorems and Proofs

In this section we present the theorems and proofs. The fact that the superactivated quantum capacity can be described by the joint output states of $\mathcal{N}_1 \otimes \mathcal{N}_2$ is summarized in Theorem 1.

**Theorem 1.** *The superactivation of joint structure $\mathcal{N}_{12} = \mathcal{N}_1 \otimes \mathcal{N}_2$ can be analyzed by the joint average $\sigma_{12}^{AB-AE}$ and joint optimal states $\rho_{12}^{AB-AE}$.*

*Proof.* Here, we show that the difference of the quantum relative entropic quantities in (14) can be positive if and only if the channels in $\mathcal{N}_{12} = \mathcal{N}_1 \otimes \mathcal{N}_2$ can activate each other, i.e., the joint channel structure is superactive. According to (14), the quantum relative entropic distance between the $\sigma_{12}^{AB-AE}$ joint average and the optimal joint state $\rho_{12}^{AB-AE}$ is equal to the

$Q^{(1)}(\mathcal{N}_1 \otimes \mathcal{N}_2)$ joint single-use quantum capacity of $\mathcal{N}_1 \otimes \mathcal{N}_2$. $Q^{(1)}(\mathcal{N}_1 \otimes \mathcal{N}_2)$ will not be superactivated if the average output joint state $\sigma_{12}^{AB-AE}$ can be given as a product state $\sigma_{12}^{AB-AE} = \sigma_1^{AB-AE} \otimes \sigma_2^{AB-AE}$. It also must find the optimal output state $\rho_{12}^{AB-AE}$, which can be given as a product state $\rho_{12}^{AB-AE} = \rho_1^{AB-AE} \otimes \rho_2^{AB-AE}$, the $Q^{(1)}(\mathcal{N}_1 \otimes \mathcal{N}_2)$. In other words, if $\sigma_{12}^{AB-AE}$ and $\rho_{12}^{AB-AE}$ can be given in a product state formula (i.e., these states are decomposable), and $Q^{(1)}(\mathcal{N}_1 \otimes \mathcal{N}_2)$ will be zero and the joint structure $\mathcal{N}_{12} = \mathcal{N}_1 \otimes \mathcal{N}_2$ will not be superactive. If these two states cannot be given in tensor product representations, then strict additivity of individual quantum capacities $Q^{(1)}(\mathcal{N}_1)$ and $Q^{(1)}(\mathcal{N}_2)$ will fail and the channel construction $\mathcal{N}_{12} = \mathcal{N}_1 \otimes \mathcal{N}_2$ will be superactive, which leads to $Q^{(1)}(\mathcal{N}_1 \otimes \mathcal{N}_2) > 0$. If the joint states $\sigma_{12}^{AB-AE}$ and $\rho_{12}^{AB-AE}$ are product states, then $Q^{(1)}(\mathcal{N}_1) = Q^{(1)}(\mathcal{N}_2) = Q^{(1)}(\mathcal{N}_1 \otimes \mathcal{N}_2) = 0$, which concludes the proof of Theorem 1.

∎

These results from the superactivation of the joint structure $\mathcal{N}_1 \otimes \mathcal{N}_2$ are extended to the properties of the joint optimal and average states in Theorem 2.

**Theorem 2.** *The quantum channels $\mathcal{N}_1$ and $\mathcal{N}_2$ of the joint structure $\mathcal{N}_{12}$ are superactive if and only if the $\sigma_{12}^{AB-AE}$ joint average state and the $\rho_{12}^{AB-AE}$ joint optimal output state of the joint channel structure are entangled states.*

*Proof.* Using the results derived by Cortese [14-15], and Petz et al. [18-20] and Schumacher and Westmoreland [16-17] the following statements can be made. The "product state formula" form expresses that the channels $\mathcal{N}_1$ and $\mathcal{N}_2$ of the joint structure $\mathcal{N}_{12}$ cannot activate each other. We use the *minimax* criterion for the joint states $\rho_{12}^{AB-AE}$ and $\sigma_{12}^{AB-AE}$ along with (14). If the joint average state and the joint optimal output state are entangled states, then the joint channel structure $\mathcal{N}_{12}$ is superactive and the quantum relative entropic distance between the joint states

$\rho_{12}^{AB-AE}$ and $\sigma_{12}^{AB-AE}$ is greater than zero. If the quantum channels $\mathcal{N}_1$ and $\mathcal{N}_2$ of the joint structure $\mathcal{N}_{12}$ can activate each other, then the informational distance of $\rho_{12}^{AB-AE}$ and $\sigma_{12}^{AB-AE}$ cannot be decomposed in the expression of the quantum relative entropy function [14-15], see (14). We will use again that quantum capacity can be expressed from the Holevo information.

If joint states $\rho_{12}^{AB-AE}$ and $\sigma_{12}^{AB-AE}$ of the joint channel $\mathcal{N}_1 \otimes \mathcal{N}_2$ are *product* states, i.e., $\rho_{12}^{AB-AE} = \rho_1^{AB-AE} \otimes \rho_2^{AB-AE}$ and $\sigma_{12}^{AB-AE} = \sigma_1^{AB-AE} \otimes \sigma_2^{AB-AE}$, then the $Q^{(1)}(\mathcal{N}_1 \otimes \mathcal{N}_2)$ joint capacity will be zero, since the quantum relative entropy function $D(\cdot \| \cdot)$ in (14) can be factorized as follows:

$$
\begin{aligned}
&Q^{(1)}(\mathcal{N}_1 \otimes \mathcal{N}_2) \\
&= \min_{\sigma_{12}} \max_{\rho_{12}} D\left(\rho_{12}^{AB} \middle\| \sigma_{12}^{AB}\right) - \min_{\sigma_{12}} \max_{\rho_{12}} D\left(\rho_{12}^{AE} \middle\| \sigma_{12}^{AE}\right) \\
&= \min_{\sigma_{12}^{AB-AE}} \max_{\rho_{12}^{AB-AE}} D\left(\rho_{12}^{AB-AE} \middle\| \sigma_{12}^{AB-AE}\right) \\
&= \min_{\sigma_{12}^{AB-AE}} \max_{\rho_{12}^{AB-AE}} Tr_{12}\left(\left(\rho_{12}^{AB-AE}\right)\log\left(\rho_{12}^{AB-AE}\right) - \left(\rho_{12}^{AB-AE}\right)\log\left(\sigma_{12}^{AB-AE}\right)\right) \\
&= \min_{\sigma_1^{AB-AE}} \min_{\sigma_2^{AB-AE}} \max_{\rho_1^{AB-AE}} \max_{\rho_2^{AB-AE}} Tr_{12}\begin{bmatrix}\left(\rho_1^{AB-AE} \otimes \rho_2^{AB-AE}\right)\log\left(\left(\rho_1^{AB-AE}\right) \otimes \left(\rho_2^{AB-AE}\right)\right) \\ -\left(\left(\rho_1^{AB-AE}\right) \otimes \left(\rho_2^{AB-AE}\right)\right)\log\left(\sigma_1^{AB-AE} \otimes \sigma_2^{AB-AE}\right)\end{bmatrix} \\
&= \min_{\sigma_1^{AB-AE}} \min_{\sigma_2^{AB-AE}} \max_{\rho_1^{AB-AE}} \max_{\rho_2^{AB-AE}} Tr_{12}\begin{bmatrix}\left(\rho_1^{AB-AE} \otimes \rho_2^{AB-AE}\right)\left(\log\left(\rho_1^{AB-AE}\right) \otimes I_2\right) + \\ \left(\rho_1^{AB-AE} \otimes \rho_2^{AB-AE}\right)\left(I_1 \otimes \log\left(\rho_2^{AB-AE}\right)\right)\end{bmatrix} \\
&\quad - \min_{\sigma_1^{AB-AE}} \min_{\sigma_2^{AB-AE}} \max_{\rho_1^{AB-AE}} \max_{\rho_2^{AB-AE}} Tr_{12}\begin{bmatrix}\left(\rho_1^{AB-AE} \otimes \rho_2^{AB-AE}\right)\left(\log\left(\sigma_1^{AB-AE}\right) \otimes I_2\right) + \\ \left(\rho_1^{AB-AE} \otimes \rho_2^{AB-AE}\right)\left(I_1 \otimes \log\left(\sigma_2^{AB-AE}\right)\right)\end{bmatrix} \\
&= \min_{\sigma_1^{AB-AE}} \min_{\sigma_2^{AB-AE}} \max_{\rho_1^{AB-AE}} \max_{\rho_2^{AB-AE}} Tr_1\left(\left(\rho_1^{AB-AE}\right)\log\left(\rho_1^{AB-AE}\right)\right)Tr_2\left(\left(\rho_2^{AB-AE}\right)I_2\right) \\
&\quad + \min_{\sigma_1^{AB-AE}} \min_{\sigma_2^{AB-AE}} \max_{\rho_1^{AB-AE}} \max_{\rho_2^{AB-AE}} Tr_1\left(\left(\rho_1^{AB-AE}\right)I_1\right)Tr_2\left(\left(\rho_2^{AB-AE}\right)\log\left(\rho_2^{AB-AE}\right)\right) \\
&\quad - \min_{\sigma_1^{AB-AE}} \min_{\sigma_2^{AB-AE}} \max_{\rho_1^{AB-AE}} \max_{\rho_2^{AB-AE}} Tr_1\left(\left(\rho_1^{AB-AE}\right)\log\left(\sigma_1^{AB-AE}\right)\right)Tr_2\left(\left(\rho_2^{AB-AE}\right)I_2\right) \\
&\quad - \min_{\sigma_1^{AB-AE}} \min_{\sigma_2^{AB-AE}} \max_{\rho_1^{AB-AE}} \max_{\rho_2^{AB-AE}} Tr_1\left(\left(\rho_1^{AB-AE}\right)I_1\right)Tr_2\left(\left(\rho_2^{AB-AE}\right)\log\left(\sigma_2^{AB-AE}\right)\right) \\
&= \min_{\sigma_1^{AB-AE}} \min_{\sigma_2^{AB-AE}} \max_{\rho_1^{AB-AE}} \max_{\rho_2^{AB-AE}} \left(Tr_1\left(\left(\rho_1^{AB-AE}\right)\log\left(\rho_1^{AB-AE}\right)\right) - Tr_1\left(\left(\rho_1^{AB-AE}\right)\log\left(\sigma_1^{AB-AE}\right)\right)\right) \\
&\quad + \min_{\sigma_1^{AB-AE}} \min_{\sigma_2^{AB-AE}} \max_{\rho_1^{AB-AE}} \max_{\rho_2^{AB-AE}} \left(Tr_2\left(\left(\rho_2^{AB-AE}\right)\log\left(\rho_2^{AB-AE}\right)\right) - Tr_2\left(\left(\rho_2^{AB-AE}\right)\log\left(\sigma_2^{AB-AE}\right)\right)\right)
\end{aligned} \quad (15)
$$

$$\begin{aligned}
&= \min_{\sigma_1^{AB-AE}} \min_{\sigma_2^{AB-AE}} \max_{\rho_1^{AB-AE}} \max_{\rho_2^{AB-AE}} \left( D\left(\rho_1^{AB-AE} \left\| \sigma_1^{AB-AE} \right.\right) + D\left(\rho_2^{AB-AE} \left\| \sigma_2^{AB-AE} \right.\right) \right) \\
&= \min_{\sigma_1^{AB-AE}} \min_{\sigma_2^{AB-AE}} \max_{\rho_1^{AB-AE}} \max_{\rho_2^{AB-AE}} D\left(\rho_1^{AB-AE} \left\| \sigma_1^{AB-AE} \right.\right) + \min_{\sigma_1^{AB-AE}} \min_{\sigma_2^{AB-AE}} \max_{\rho_1^{AB-AE}} \max_{\rho_2^{AB-AE}} D\left(\rho_2^{AB-AE} \left\| \sigma_2^{AB-AE} \right.\right) \\
&= \min_{\sigma_1^{AB-AE}} \max_{\rho_1^{AB-AE}} D\left(\rho_1^{AB-AE} \left\| \sigma_1^{AB-AE} \right.\right) + \min_{\sigma_2^{AB-AE}} \max_{\rho_2^{AB-AE}} D\left(\rho_2^{AB-AE} \left\| \sigma_2^{AB-AE} \right.\right) \\
&= Q^{(1)}\left(\mathcal{N}_1 \otimes \mathcal{N}_2\right) = Q^{(1)}\left(\mathcal{N}_1\right) + Q^{(1)}\left(\mathcal{N}_2\right) = 0,
\end{aligned}$$

where $I_1$ and $I_2$ are the $d$ dimensional identity matrices ($d=2$ for the qubit case), $\rho_{12}^{AB}$ is the optimal output state of the joint channel $\mathcal{N}_{12}$ between Alice and Bob, and $\sigma_{12}^{AB} = \sum_i p_i \rho_{12}^{AB(i)}$ is the average state of the joint channel $\mathcal{N}_{12}$ between Alice and Bob. The term $E$ denotes the environment, $\rho_{12}^{AE}$ is the optimal state of the channel between Alice and the environment, $\sigma_{12}^{AE} = \sum_i p_i \rho_{12}^{AE(i)}$ is the average state of the channel between Alice and the environment, $\rho_{12}^{AB-AE}$ is the final optimal output channel state, and $\sigma_{12}^{AB-AE}$ is the final output average state of the joint channel $\mathcal{N}_1 \otimes \mathcal{N}_2$.

The factorization of (14) implies that the single-use joint quantum capacity $Q^{(1)}\left(\mathcal{N}_1 \otimes \mathcal{N}_2\right)$ can be derived from the strict sum of independent channel quantum capacities $Q^{(1)}\left(\mathcal{N}_1\right)$ and $Q^{(1)}\left(\mathcal{N}_2\right)$, thus $Q^{(1)}\left(\mathcal{N}_1\right) = Q^{(1)}\left(\mathcal{N}_2\right) = Q^{(1)}\left(\mathcal{N}_{12}\right) = 0$. If the quantum relative entropic distance of the $\sigma_{12}^{AB-AE}$ joint average and $\rho_{12}^{AB-AE}$ joint optimal states of $\mathcal{N}_1 \otimes \mathcal{N}_2$ can be factorized, then the joint states $\sigma_{12}$ and $\rho_{12}$ of the joint channel $\mathcal{N}_{12}$ cannot be entangled states; the superactivation of the joint channel structure $\mathcal{N}_{12}$ is possible if and only if the joint states $\rho_{12}^{AB-AE}$ and $\sigma_{12}^{AB-AE}$ of the joint channel $\mathcal{N}_{12}$ are entangled states. The result on the asymptotic quantum capacity of the joint channel $\mathcal{N}_1 \otimes \mathcal{N}_2$ is

$$\begin{aligned}
&Q(\mathcal{N}_1 \otimes \mathcal{N}_2) \\
&= \lim_{n\to\infty} \frac{1}{n} \sum_n \left( \min_{\sigma_1^{AB-AE}} \min_{\sigma_2^{AB-AE}} \max_{\rho_1^{AB-AE}} \max_{\rho_2^{AB-AE}} D\left(\rho_1^{AB-AE} \middle\| \sigma_1^{AB-AE}\right) \right) \\
&\quad + \lim_{n\to\infty} \frac{1}{n} \sum_n \left( \min_{\sigma_1^{AB-AE}} \min_{\sigma_2^{AB-AE}} \max_{\rho_1^{AB-AE}} \max_{\rho_2^{AB-AE}} D\left(\rho_2^{AB-AE} \middle\| \sigma_2^{AB-AE}\right) \right) \\
&= \lim_{n\to\infty} \frac{1}{n} \sum_n \left( \min_{\sigma_1^{AB-AE}} \max_{\rho_1^{AB-AE}} D\left(\rho_1^{AB-AE} \middle\| \sigma_1^{AB-AE}\right) \right) \\
&\quad + \lim_{n\to\infty} \frac{1}{n} \sum_n \left( \min_{\sigma_2^{AB-AE}} \max_{\rho_2^{AB-AE}} D\left(\rho_2^{AB-AE} \middle\| \sigma_2^{AB-AE}\right) \right) \\
&= Q(\mathcal{N}_1 \otimes \mathcal{N}_2) = Q(\mathcal{N}_1) + Q(\mathcal{N}_2) = 0.
\end{aligned} \quad (16)$$

These results conclude the proof of Theorem 2.

∎

From Theorem 2 also follows that possible set of superactive of quantum channels $\mathcal{N}_1 \otimes \mathcal{N}_2$ is also limited by the mathematical properties of the quantum relative entropy function. This result is extended to the superactivation of *arthitrary*[†] *channel capacities* in Theorems 3 and 4.

**Theorem 3.** *The superactivation of any[†] channel capacities of the joint structure $\mathcal{N}_1 \otimes \mathcal{N}_2$ is determined by the properties of the quantum relative entropy function.*

*Proof.* The main results already are shown in (15); however, further statements can be derived from these decompositions. Factoring relative entropy function $D(\cdot\|\cdot)$ in (14) does not work if the quantum channels in $\mathcal{N}_{12}$ can activate each other, thus for entangled joint states $\rho_{12}$ and $\sigma_{12}$, the strict channel additivity will not hold for the zero-capacity channels $\mathcal{N}_1$ and $\mathcal{N}_2$. In that case, the joint channel $\mathcal{N}_1 \otimes \mathcal{N}_2$ is superactive, and the joint capacity of $\mathcal{N}_1 \otimes \mathcal{N}_2$ will be positive. If the average output state $\sigma_{12}$ of $\mathcal{N}_1 \otimes \mathcal{N}_2$ is a product state, and if one or more from the set of optimal joint output states $\rho_{12}$ is a product state, then the factorization of the quantum relative entropy function $D(\cdot\|\cdot)$ indicates that the quantum channels $\mathcal{N}_1$ and $\mathcal{N}_2$ cannot activate each other. These results along with the proof of Theorem 2 conclude the proof of Theorem 3.

---

[†] *Any channel capacities of quantum channels, for which the superactivation of the joint channel structure is theoretically possible. These capacities are: quantum capacity (single-use and asymptotic), classical and quantum zero-error capacities (single-use and asymptotic) of quantum channels.*

∎

**Theorem 4.** *The superactivation of the any[†] possible channel capacities of the joint channel construction $\mathcal{N}_1 \otimes \mathcal{N}_2$ is possible if and only if the quantum relative entropic distance of the $\sigma_{12}$ joint average and $\rho_{12}$ joint optimal states cannot be factorized.*

*Proof.* This result is follows from our previously derived result on the factorization of the quantum relative entropy function. If the joint channel structure $\mathcal{N}_1 \otimes \mathcal{N}_2$ is superactive, then the quantum relative entropic distance of the $\sigma_{12}$ joint average and $\rho_{12}$ joint optimal states cannot be factorized. Using the theories of the paper, if the channel combination $\mathcal{N}_1 \otimes \mathcal{N}_2$ is not superactive, (i.e., strict additivity holds) then the factorization of the quantum relative entropy function can be made, and the quantum relative entropic distance of the $\sigma_{12}$ joint average and $\rho_{12}$ can be expressed as the strict sum of the quantum relative entropic distances between the states $\sigma_1, \rho_1$ and $\sigma_2, \rho_2$ of the two "not superactivated" quantum channels $\mathcal{N}_1$ and $\mathcal{N}_2$. The superactivation of $\mathcal{N}_1 \otimes \mathcal{N}_2$ hold if and only if the quantum relative entropic distance between the $\sigma_{12}$ joint average state and $\rho_{12}$ joint optimal states is not decomposable to the sum of the two quantum relative entropy functions $D(\cdot\|\cdot)$ between the density matrices $\sigma_1, \rho_1$ and $\sigma_2, \rho_2$, and it also remain true in the asymptotic setting for $n \to \infty$.

∎

## 4 Conclusions

In this paper we proved that the properties of the quantum relative function also determine the superactivation of quantum channels. Our purely mathematical results have demonstrated that the effect of superactivation also depends not only on the channel maps and the properties of the quantum channels of the joint structure as was known before, but on the basic properties of the quantum relative entropy function. Before our work this connection was completely unrevealed in Quantum Information Theory.

## Acknowledgements

The results discussed above are supported by the grant TAMOP-4.2.2.B-10/1--2010-0009 and COST Action MP1006.

## References

[1] S. Imre, L. Gyongyosi: *Advanced Quantum Communications: An Engineering Approach*, Wiley-IEEE Press, (2012).

[2] S. Imre, F. Balázs: *Quantum Computing and Communications – An Engineering Approach*, Published by John Wiley and Sons Ltd, (2005).

[3] M. Hastings, "A Counterexample to Additivity of Minimum Output Entropy" *Nature Physics* 5, 255, arXiv:0809.3972, (2009).

[4] G. Smith, J. Yard, Quantum Communication with Zero-capacity Channels. *Science* 321, 1812-1815 (2008).

[5] G. Smith, J. A. Smolin and J. Yard, Quantum communication with Gaussian channels of zero quantum capacity, *Nature Photonics* 5, 624–627 (2011), arXiv:1102.4580v1, (2011).

[6] R. Duan, Superactivation of zero-error capacity of noisy quantum channels.arXiv:0906.2527, (2009).

[7] T. Cubitt, J. X. Chen, and A. Harrow, Superactivation of the Asymptotic Zero-Error Classical Capacity of a Quantum Channel, *IEEE Trans. Inform. Theory* 57, 8114 (2011).arXiv: 0906.2547. (2009).

[8] T. Cubitt, G. Smith, An Extreme form of Superactivation for Quantum Zero-Error Capacities, *IEEE Trans. Inf. Theory* 58, 1953 (2012). arXiv:0912.2737v1.

[9] S. Lloyd, "Capacity of the noisy quantum channel," *Phys. Rev. A*, vol. 55, pp. 1613–1622, (1997)

[10] P. Shor, "The quantum channel capacity and coherent information." lecture notes, MSRI Workshop on Quantum Computation, Available online at http://www.msri.org/publications/ln/msri/2002/quantumcrypto/shor/1/. (2002).

[11] I. Devetak, "The private classical capacity and quantum capacity of a quantum channel," *IEEE Trans. Inf. Theory,* vol. 51, pp. 44–55, quant-ph/0304127, (2005).

[12] A. Holevo, "The capacity of the quantum channel with general signal states", *IEEE Trans. Info. Theory* 44, 269 - 273 (1998).

[13] B. Schumacher and M. Westmoreland, "Sending classical information via noisy quantum channels," *Phys. Rev. A*, vol. 56, no. 1, pp. 131–138, (1997).

[14] J. Cortese, "The Holevo-Schumacher-Westmoreland Channel Capacity for a Class of Qudit Unital Channels", LANL ArXiV e-print quant-ph/0211093, (2002).

[15] J. Cortese, "*Classical Communication over* Quantum *Channels*". PhD Thesis by. John A. Cortese. California Institute of Technology (2003).

[16] B. Schumacher and M. Westmoreland, "Optimal Signal Ensembles", LANL ArXiV e-print quant-ph/9912122, (1999).

[17] B. Schumacher and M. Westmoreland, "Relative Entropy in Quantum Information Theory" 2000, LANL ArXiV e-print quant-ph/0004045, to appear in *Quantum Computation and Quantum Information: A Millenium Volume* , S. Lomonaco, editor (American Mathematical Society Contemporary Mathematics series), (2000).

[18] D. Petz and C. Sudár, "Geometries of quantum states," *Journal of Mathematical Physics*, vol. 37, no. 6, pp. 2662–2673, (1996).

[19] D. Petz, Bregman divergence as relative operator entropy, Acta Math. Hungar, 116, 127-131. (2007).

[20] D. Petz, *Quantum Information Theory and Quantum Statistics*: Springer-Verlag, Heidelberg, Hiv: 6. (2008).

[21] L. Gyongyosi, S. Imre: Algorithmic Superactivation of Asymptotic Quantum Capacity of Zero-Capacity Quantum Channels, *Information Sciences*, ELSEVIER, ISSN: 0020-0255; 2011.

[22] L. Gyongyosi, S. Imre: Classical Communication with Stimulated Emission over Zero-Capacity Optical Quantum Channels, *APS DAMOP 2012 Meeting,* The 43rd Annual Meeting of the APS Division of Atomic, Molecular, and Optical Physics, (American Physical Society), Jun. 2012, Anaheim, California, USA.

[23] L. Hanzo, H. Haas, S. Imre, D. O'Brien, M. Rupp, L. Gyongyosi: Wireless Myths, Realities, and Futures: From 3G/4G to Optical and Quantum Wireless, *Proceedings of the IEEE*, Volume: 100 , Issue: Special Centennial Issue, pp. 1853-1888.

[24] F. Brandao and J. Oppenheim, "Public Quantum Communication and Superactivation," arXiv:1005.1975. (2010).

[25] F. Brandao, J. Oppenheim and S. Strelchuk, "When does noise increase the quantum capacity?", *Phys. Rev. Lett.* 108, 040501 (2012), arXiv:1107.4385v1 [quant-ph]

[26] K. Bradler, P. Hayden, D. Touchette, and M. M. Wilde, Trade-off capacities of the quantum Hadamard channels, *Journal of Mathematical Physics* 51, 072201, arXiv:1001.1732v2, (2010).

[27] L. Gyongyosi, S. Imre, On the Mathematical Boundaries of Communication with Zero-Capacity Quantum Channels, *Proceedings of the Turing-100. The Alan Turing Centenary Conference*, 2012.

[28] J. Park, S. Lee, Zero-error classical capacity of qubit channels cannot be superactivated, *Phys. Rev. A* 85, 052321 (2012), arXiv:1205.5851v1 [quant-ph] 26 May 2012.